\documentclass{jqsrt} 

\setpg{1} 

\titl[Polarized Scattering and Biosignatures in Exoplanetary Atmospheres]
{Polarized Scattering and Biosignatures in Exoplanetary Atmospheres} 

\authors{S.V.\,Berdyugina}{S.V.\,Berdyugina\aff{1}}

\affiliat{
 \item Kiepenheuer Institut f\"ur Sonnenphysik, Freiburg, Germany
}

\email{berdyugina@kis.uni-freiburg.de} 

\begin{document} 

\begin{abstract}
  
Polarized scattering in planetary atmospheres is computed in the context of exoplanets.
The problem of polarized radiative transfer is solved for a general case of absorption and scattering,
while Rayleigh and Mie polarized scattering are considered as most relevant examples.
We show that 
(1) relative contributions of single and multiple scattering depend on the stellar irradiation 
and opacities in the planetary atmosphere; 
(2) cloud (particle) physical parameters can be deduced from the wavelength-dependent
measurements of the continuum polarization and from a differential analysis of molecular band absorption; 
(3) polarized scattering in molecular bands increases the reliability of their detections in exoplanets;
(4) photosynthetic life can be detected on other planets in visible polarized spectra with high sensitivity.
These examples demonstrate the power of spectropolarimetry for exoplanetary research and
for searching for life in the universe.
  
\end{abstract}  

\section{Polarized Radiative Transfer}\label{sec:prt}

Radiative processes in planetary atmospheres is a classical subject,
simply for the reason that we live in one. Extensive theoretical studies 
were carried out during the second half of the twentieth century by such giants as 
Sobolev \cite{Sobolev1956} and Chandrasekhar \cite{Chandra1960} 
as well as the renown radiative transfer school at the Saint Petersburg (Leningrad)
University \cite{Nagirner2016}. Most recently, physics of planetary atmospheres has become 
one of the most acclaimed subjects because of applications for Earth climate studies and
the detection of a large variety of extrasolar planets. This paper provides the theoretical
basis for studying atmospheres of exoplanets using techniques of spectropolarimetry available to us. 
In particular, 
using molecular band and continuum spectropolarimetry one can reveal the composition
of the gaseous atmosphere, particle layers (clouds, hazes, etc.) and the planetary surface,
including the land, water, and life. Modeling these cases is described in this paper.

We start from solving a self-consistent radiative transfer problem for polarized scattering in a planetary
atmosphere illuminated by a host star. We solve this problem under the following assumptions:
	 1) the atmosphere is plane-parallel and static;
	 2) the planet is spherically symmetric;
	 3) stellar radiation can enter the planetary atmosphere from different angles and can be polarized;
	 4) an incoming photon is either absorbed or scattered according to opacities in the atmosphere;
	 5) an absorbed photon does not alter the atmosphere (model atmosphere includes thermodynamics effects 
	      of irradiation);
	 6) photons can be scattered multiple times until they escape the atmosphere.
These assumptions expand those in \cite{FluriBerd2010}, namely that multiple scattering 
is allowed, stellar irradiation can be polarized and vary with an incident angle, and the planetary
atmosphere can be inhomogeneous in both longitude and latitude.

Then, the radiative transfer equation for the Stokes vector ${\bf I}=(I,Q,U,V)^\mathrm{T}$ 
of scattered polarized radiation of a given frequency (omitted for clarity) towards ($\mu=\cos\theta$, $\varphi$) is
\begin{equation}\label{eq:rt}
\mu\frac{d\bf{I}(\tau,\mu,\varphi)}{d\tau} = \bf{I}(\tau,\mu,\varphi) - \bf{S}(\tau,\mu,\varphi)
\end{equation}
with the total source function
\begin{equation}\label{eq:sourcef}
\bf{S}(\tau,\mu,\varphi) = \frac{\kappa(\tau)\bf{B}(\tau)+\sigma(\tau)\bf{S}_\mathrm{sc}(\tau,\mu,\varphi)}{\kappa(\tau)+\sigma(\tau)}
\rm \ ,
\end{equation}
where $\kappa$ and $\sigma$ are absorption and scattering opacities, 
$\bf{S}_\mathrm{sc}$ and $\bf{B}$  are the scattering source function and 
the unpolarized thermal emission, respectively, and $\tau$ is the optical depth in the atmosphere
with $\tau=0$ at the top. The formal solution of Eq.~(\ref{eq:rt}) is (e.g., \cite{Sobolev1956})
\begin{equation}\label{eq:rt_formal}
\bf{I}(\tau,\mu,\varphi) = \bf{I}(\tau_*,\mu,\varphi)e^{-(\tau_*-\tau)/\mu}
                 + \int_{\tau}^{\tau_*} \bf{S}(\tau',\mu,\varphi)e^{-(\tau'-\tau)/\mu}\frac{\mathrm{d}\tau'}{\mu}
\rm \ ,
\end{equation}
where $\tau_*$ is either the optical depth at the bottom of the atmosphere for the Stokes vector 
$\bf{I}^+(\tau,\mu,\varphi)$ coming from the bottom to the top ($\theta < \pi/2$) or the optical depth 
at the top of the atmosphere ($\tau_*=0$) for the Stokes vector $\bf{I}^-(\tau,\mu,\varphi)$ coming from 
the top to the bottom ($\theta > \pi/2$).

The scattering source function $\bf{S}_\mathrm{sc}$ is expressed via the scattering phase matrix 
$\bf{\hat{P}}(\mu,\mu';\varphi,\varphi')$, depending on the directions of the incident ($\mu'$, $\varphi'$) 
and scattered ($\mu$, $\varphi$) light:
\begin{equation}\label{eq:sourcef_sc}
\bf{S}_\mathrm{sc}(\tau,\mu,\varphi) = \int \bf{\hat{P}}(\mu,\mu';\varphi,\varphi')\bf{I}(\tau,\mu',\varphi')\frac{d\Omega'}{4\pi}
\rm \ .
\end{equation}
It has contributions from scattering both incident stellar light and intrinsic thermal emission. 
Their relative contributions depend on the frequency. For instance, for Rayleigh scattering 
the intensity of the thermal emission of a relatively cold planet in the blue part of the spectrum 
may become negligible compared to that of the scattered stellar light. The phase matrix 
$\bf{\hat{P}}(\mu,\mu';\varphi,\varphi')$ is a $4\times4$ matrix with six independent parameters for scattering cases 
on particles with a symmetry \cite{HansenTravis1974}. In this paper we employ the Rayleigh 
and Mie scattering phase matrices but our formalism is valid for other phase functions too.

The Stokes vector of the light emerging from the planetary atmosphere ${\bf I}(0,\mu,\varphi)$ is obtained 
by integrating iteratively Equations (\ref{eq:sourcef}) and (\ref{eq:rt_formal}) for a given vertical 
distribution of the temperature and opacity in a planetary atmosphere. Boundary conditions are defined 
by stellar irradiation at the top, planetary thermal radiation at the bottom, 
and (if present) reflection from the planetary surface. Stellar irradiation can be polarized, 
but the planetary thermal radiation is unpolarized. 
In particular, stellar limb darkening and linear polarization due to scattering in the stellar 
atmosphere \cite{KostoBerd2015,Kosto2016} can be taken into account, including the influence
of dark spots on the stellar surface \cite{Kosto2015}. This effect is not very large but may be
important for cooler stars with large spots and planets on very short-period orbits (when the stellar
radiation incident angle noticeably vary depending on the stellar limb angle). 
Also, stellar magnetic fields causing polarization in stellar line profiles due to the Zeeman 
effect can be included for given atomic and molecular lines \cite{Berd2003}. This effect is only 
important for high-resolution spectropolarimetry which is not yet possible for exoplanets.
Depending on the structure of the phase matrix and the boundary conditions, the equations are solved 
for all or a fewer Stokes vector components. Normally it takes 3--7 
iterations to achieve a required accuracy. The radiation flux is then obtained by integrating 
the Stokes vector over the illuminated planetary surface with a coordinate grid 
($6^\circ\times6^\circ$) on the planetary surface for a given orbital phase angle as described 
in \cite{FluriBerd2010}. 

Our model includes the following opacity sources: 
(1) Rayleigh scattering on H I, H$_2$, He I, H$_2$O, CO, CH$_4$ and other relevant molecules, 
    Thomson scattering on electrons, and
    Mie scattering on spherical particles with a given size distribution, 
    with all scattering species contributing to the continuum polarization,
(2) absorption in the continuum due to free-free and bound-free transitions of 
    H I, He I, H$^-$, H$_2$$^+$, H$_2$$^-$, He$^-$, metal ionization,
	and collision-induced absorption (CIA) by H$_2$--H$_2$;
(3) absorption and scattering in atomic and molecular lines for particular frequencies where they contribute.
Number densities of the relevant species are calculated with a chemical equilibrium code described in
\cite{Berd2003}. Here we employ model atmospheres from \cite{Allard2001} and \cite{Witte2009} for 
stellar and planetary atmospheres, respectively, according to their effective temperatures (T$_{\rm eff}$). 
This is appropriate for illustrating radiative transfer effects discussed in Section~\ref{sec:res} 
and applicable for the case of highly irradiated hot Jupiters and substellar components.
In particular, a model atmosphere of a hot Jupiter has to match the infrared thermal radiation 
of the planet originating in deeper layers, while upper layers contributing to the optical radiation 
are completely dominated by the incident stellar radiation. 
Planetary atmosphere models with specific chemical compositions and temperature-pressure 
(TP) structures can be also employed. For instance, the planetary atmosphere can be inhomogeneous
with the vertical composition and TP-structure varying with latitude and longitude.

\section{Results}\label{sec:res}

\subsection{Rayleigh Scattering}\label{sec:Ray}

In this section we assume that scattering in the planetary atmospheres occurs only on
atoms, molecules or particles which are significantly smaller than the wavelength
of scattered light, i.e., we employ the Rayleigh scattering phase matrix, including
isotropic scattering intensity. 
In particular, we focus here on examples of resulting Stokes parameters and source functions
depending on stellar irradiation and wavelength.

\begin{figure}[h]
  \centering
  \includegraphics[width=0.25\paperwidth]{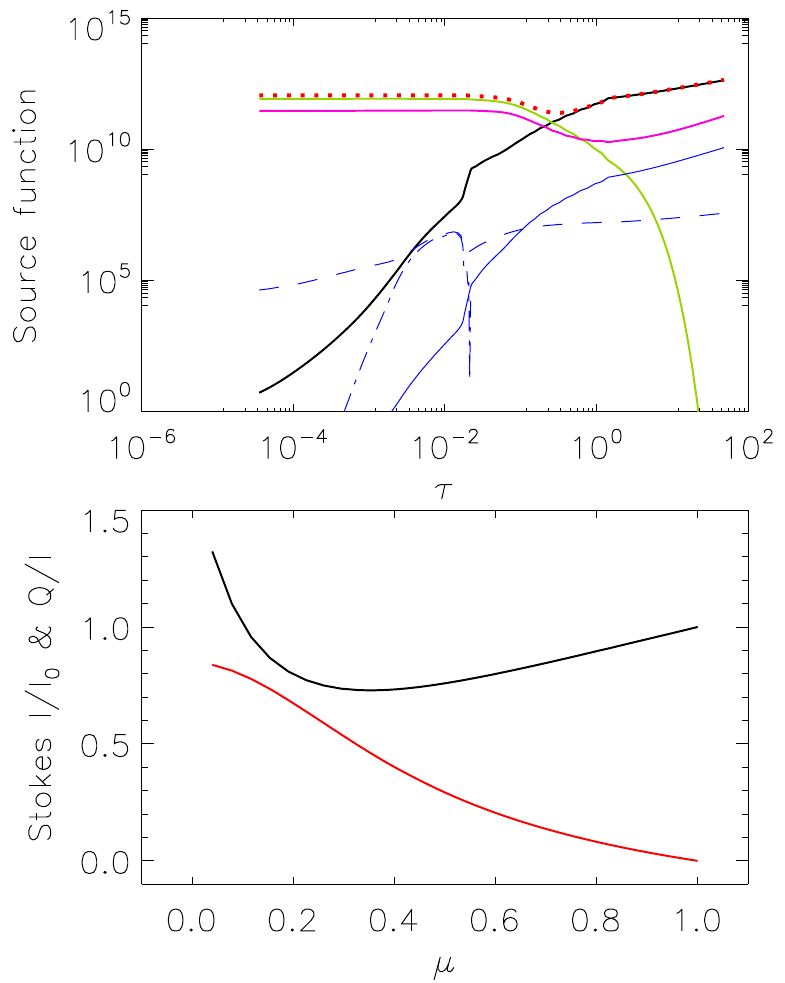}
  \includegraphics[width=0.25\paperwidth]{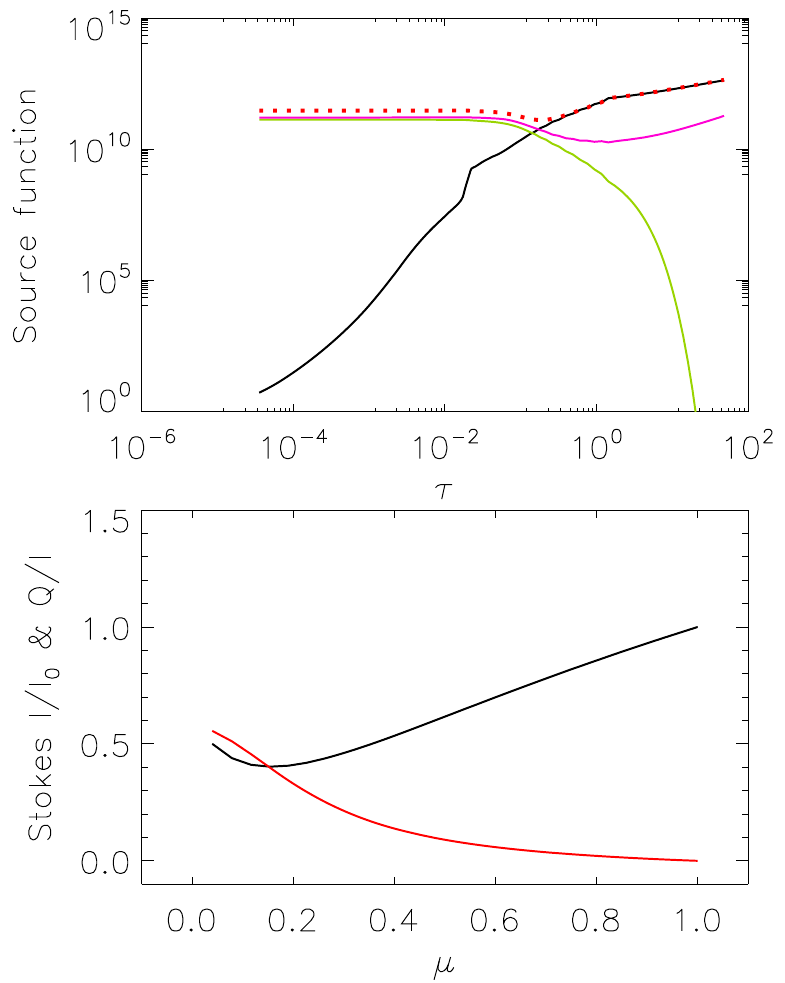}
  \includegraphics[width=0.25\paperwidth]{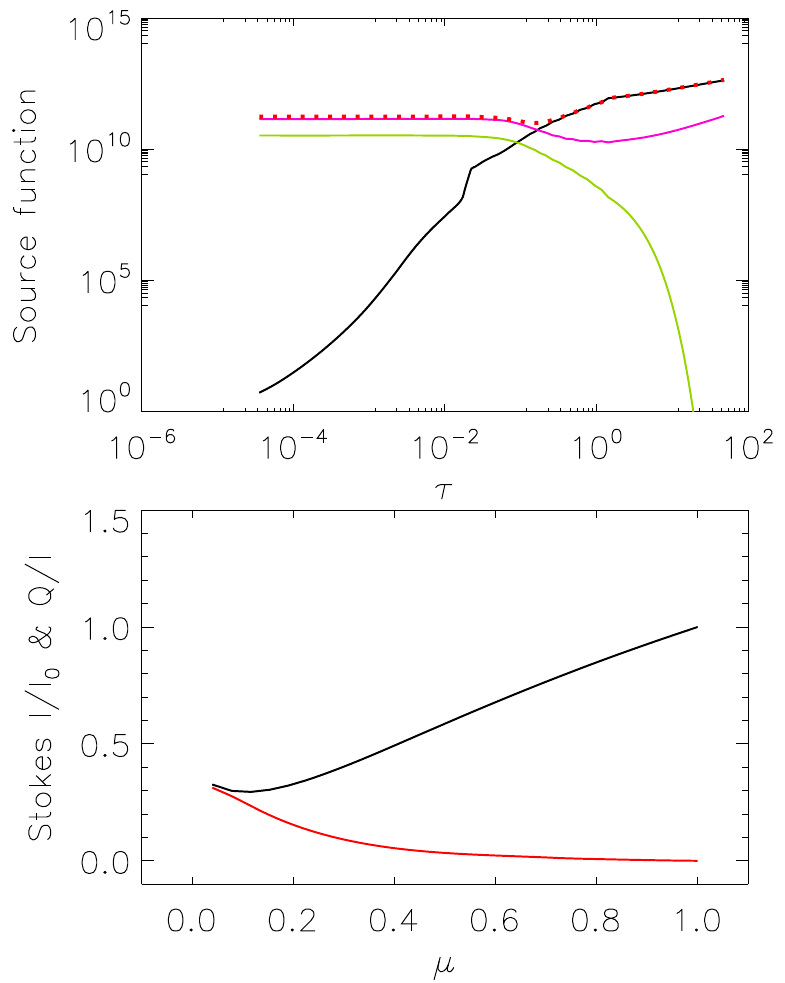}
  \caption{Stokes $I$ source functions (top panels) and normalized emerging Stokes $I$ and $Q$ parameters 
  (bottom panels) for three distances between the star and the planet: 0.02, 0.05, and 0.1 AU (left to right). 
  The source functions are shown separately for thermal radiation of the planet (black), 
  single scattered stellar radiation (green), multiple scattered stellar and planetary radiation
  (magenta), and the total one (red dotted line). The top left plot also shows relative scattering 
  (dashed blue) and absorption (solid blue) opacities and separately particle scattering (dashed-dotted blue) 
  as a cloud layer in the original model atmosphere (it is the same for all three panels). 
  In the lower panels, the Stokes $I/I_0$ (black) is normalized to the intensity
  at the planet disk centre ($\mu=0$). The Stokes $Q/I$ (red) is normalized to to $I$ at given $\mu$. Both
  are at $\tau=0$. Notice the increase of the single scattering contribution with respect to that of 
  multiple scattering as the the distance to the planet decreases (i.e., the stellar flux increases).
  Accordingly, the planet limb polarization and brightening increase too.
  }
  \label{fig:ray}
\end{figure}

Figure~\ref{fig:ray} shows examples of depth dependent Stokes $I$ source functions (top panels) and 
normalized emerging Stokes $I$ and $Q$ (bottom panels) for three distances between the star and the planet
(left to right) at the wavelength of 400nm. 
Here, the star is of T$_{\rm eff}=5500$\,K, and the planet is of T$_{\rm eff}=1500$\,K.
Stokes $Q$ is assumed to be positive when polarization is perpendicular to the scattering plane.
%
%
By studying the behaviour of the source functions and Stokes parameters depending on various parameters
we conclude the following facts:
\begin{itemize}
\item The polarization at a given depth in the atmosphere arises due to its anisotropic irradiation,
i.e., unequal illumination coming from the top and from the bottom (assuming here an azimuthal
symmetry). Hence, anisotropy and polarization are small in deeper layers, where planet thermal
radiation dominates, and they are larger in upper layers, where stellar irradiation dominates.
The depth where this dominance alternates depends on the relative contribution of the scattering 
and absorption coefficients to the total opacity (which is wavelength dependent). 
It turns out that in cool gaseous atmospheres 
this occurs very deep in the atmosphere for the continuum radiation, but can be higher for radiation 
in cores of strong absorption atomic and molecular lines. 

\item This anisotropy (and, hence, polarization) is sensitive to the incident stellar flux 
(cf., number of photons arriving to the planet) at wavelengths where Rayleigh scattering is most 
efficient, i.e., in the blue part of the spectrum. Thus, hotter stars hosting closer-in planets 
are systems potentially producing larger polarization in the blue. 

\item Relative contribution of single-scattered photons with larger polarization with respect to 
multiple-scattered photons with lower resulting polarization increases with stellar irradiation
at shorter wavelengths. 

\item Depending on stellar irradiation, intensity distribution on the planetary disk,
i.e., $I(0,\mu)/I(0,1)$ can decrease or increase with $\mu$. In fact, the $\mu$ value where
limb darkening turns into limb brightening approximates the optical depth $\tau$ where single
and multiple scattering contributions become comparable.

\item Planet limb polarization is very sensitive to the stellar irradiation because of the effects
listed above. For a larger stellar flux, a larger polarization is seen for a wider range of angles.

\item Considering the high sensitivity of planet polarization to stellar irradiation, variability 
of the stellar flux incident on the planet, e.g., caused by dark (magnetic) spots or flares, 
can result in a variable {\it amplitude} of planet polarization, while its orbital phase dependence 
is preserved, since the latter depends on orbital parameters only (see \cite{FluriBerd2010}).
\end{itemize}

The models presented in Figure~\ref{fig:ray} are close to the case of the HD189733b hot Jupiter 
which is at about 0.03 AU from its K-dwarf star with the effective temperature of about 5500\,K.
The relatively high polarization measured from this planel in the blue band (B-band) 
\cite{Berd2008,Berd2011} is well explained by the dominance of the single-scattered stellar photons
in its upper atmosphere because of the high irradition and Rayleigh scattering cross-section in the blue. 
This was first proposed in \cite{Berd2011} and further demonstrated with a simple model in \cite{BerdSPW2011}.
Here, with the precise calculations of the polarized radiative transfer, we show that this hypothesis is valid.
Moreover, multi-wavelength polarimetry allowed for estimating the planet albedo and determining its blue colour.
The relation between the geometrical albedo and polarization is however not so simple as was assumed in 
\cite{Berd2011}. An analysis of this relation for various planetary and stellar parameters using this
theory will be carried out in a separate paper.

\subsection{Mie Scattering}\label{sec:Mie}

In this section we consider scattering caused by spherical particles of various 
sizes which can be comparable or larger than the wavelength of scattered light, 
i.e., we employ the Mie scattering phase matrix.
For smaller particles and/or longer wavelengths it approximates Rayleigh scattering.

\begin{figure}[h]
  \centering
  \includegraphics[width=0.65\paperwidth]{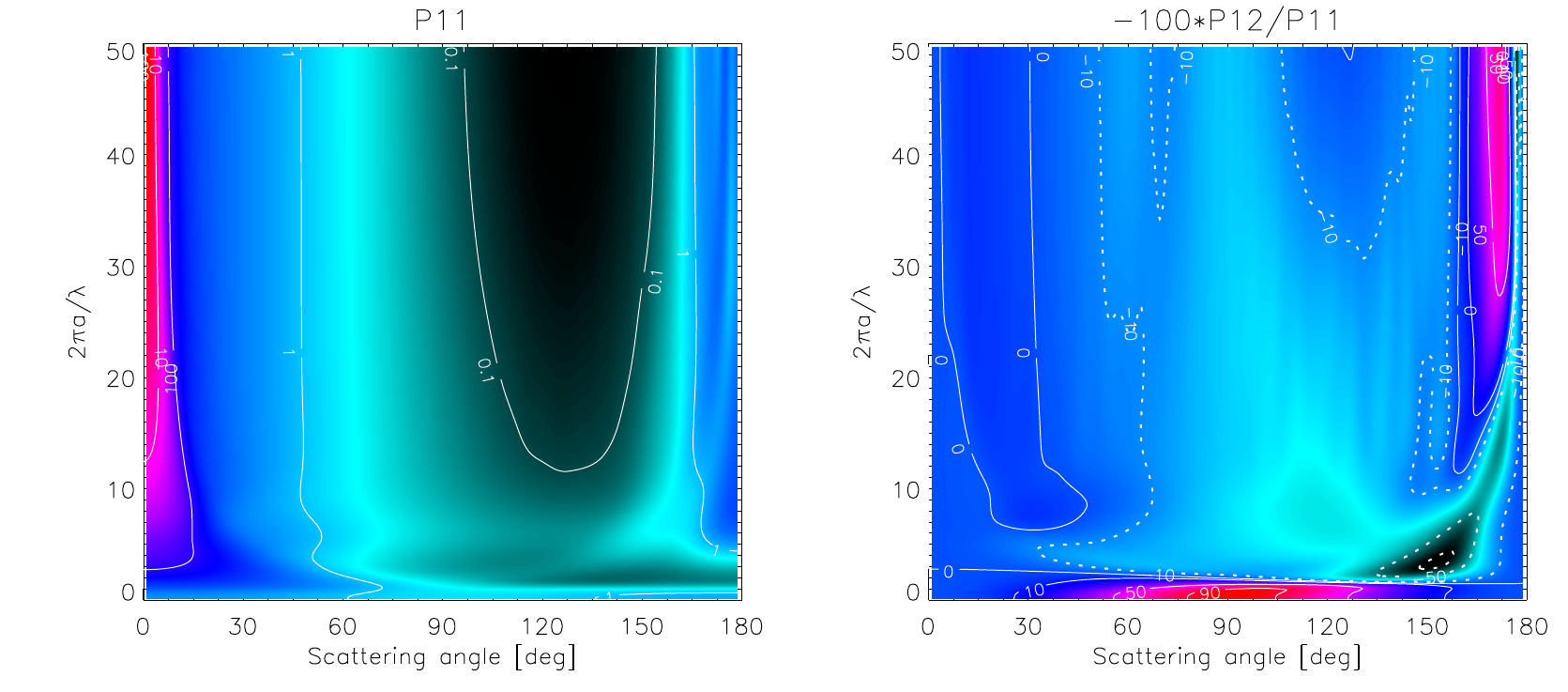}\\
  \includegraphics[width=0.65\paperwidth]{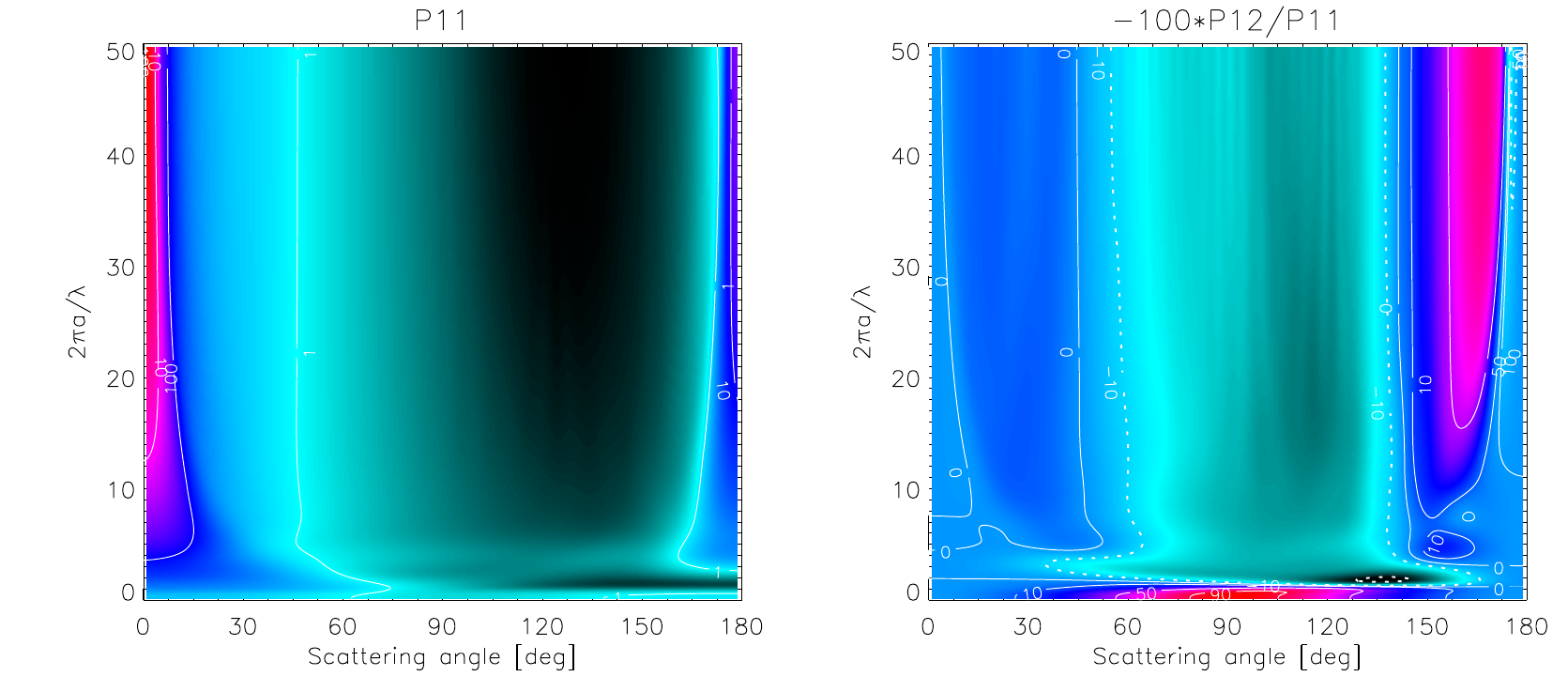}
  \caption{Mie scattering phase matrix elements $P_{11}$ (intensity) and $-100\%P_{21}/P_{11}$
  (percent polarization)  for single scattering on particles with  $n_r$=1.6 (top panels)
  and $n_r$=1.9 (bottom panels) and the effective particle size variance $b$=0.07.
  Dotted lines are contours for negative polarization (parallel to the scattering plane).}
  \label{fig:mie}
\end{figure}

Following examples in earlier literature (e.g., \cite{HansenTravis1974} and references therein),
we assume "gamma" distribution of particle sizes: $n(r)=\mathrm{C}r^{(1-3b)/b}e^{-r/ab}$, 
with $a$ and $b$ being the effective particle radius and the effective size variance, respectively.
Also, we use the so-called size parameter $2\pi a/\lambda$ which can be recalculated to $\lambda$
for a given $a$, and vice versa. Particles are characterized by the refractive index $n_r$ with 
its real part being responsible for scattering. With this, we can reproduce numerical examples 
in \cite{HansenTravis1974} as well as results for Venus in \cite{HansenHovenier1971}. 
Here, we investigate scattering on highly refractive materials 
($n_r > 1.5$) which are expected to be present in hot Jupiter atmospheres. 
For instance, olivine, which is common in the Solar system, and its endmembers forsterite 
Mg$_2$SiO$_4$ and fayalite Fe$_2$SiO$_4$ have a range of the refractive index from 1.6 to 1.9. 

In Fig.~\ref{fig:mie} we show examples of two Mie phase matrix elements: 
intensity $P_{11}$ and percent polarization $-100\%P_{21}/P_{11}$ for single scattering 
on particles with $n_r$ of 1.6 (upper panels) and 1.9 (lower panels), depending on the
size parameter and the scattering angle. The latter is 0$^\circ$ for forward
and 180$^\circ$ for backward scattering.
These examples illustrate the following known facts 
(e.g., \cite{vandeHulst1957,HansenHovenier1971,HansenTravis1974}):
\begin{itemize}
\item Forward scattering dominates the intensity for larger particles. 
\item For the smallest particle size parameters polarization is strong (up to 100\%) 
      and positive near scattering angle 90$^\circ$ due to Rayleigh scattering. For the largest size
      parameters polarization approaches that of the geometrical optics, i.e., it is small at
      small scattering angles because of largely unpolarized diffracted light, and it is negative for
      a wide range of angles because of two refractions within a sphere. 
\item Strong positive polarization maximum near 165$^\circ$--170$^\circ$ is the primary
      rainbow. It can reach 100\%\ polarization for certain size-angle combinations. 
\item Strong negative polarization near 140$^\circ$--150$^\circ$ is a "glory"-like phenomenon,
      caused by surface waves on the scattering particle. The "glory" itself, which is a sharp
	  maximum in polarization in the backscattering direction, can be seen on particles with
	  larger size parameters. 
\item Weak positive polarization near 20$^\circ$--30$^\circ$ is due to "anomalous diffraction"
      caused by interference of the diffracted, reflected and transmitted light in the forward
	  direction.
\end{itemize}

\begin{figure}[h]
  \centering
  \includegraphics[width=0.25\paperwidth]{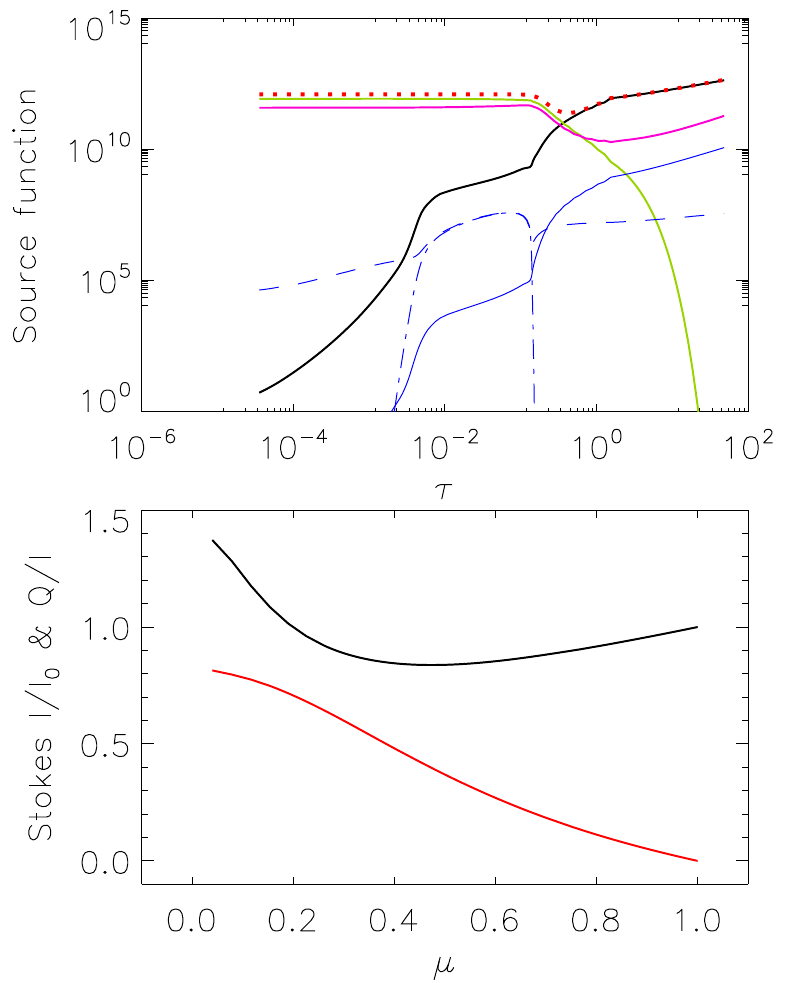}
  \includegraphics[width=0.25\paperwidth]{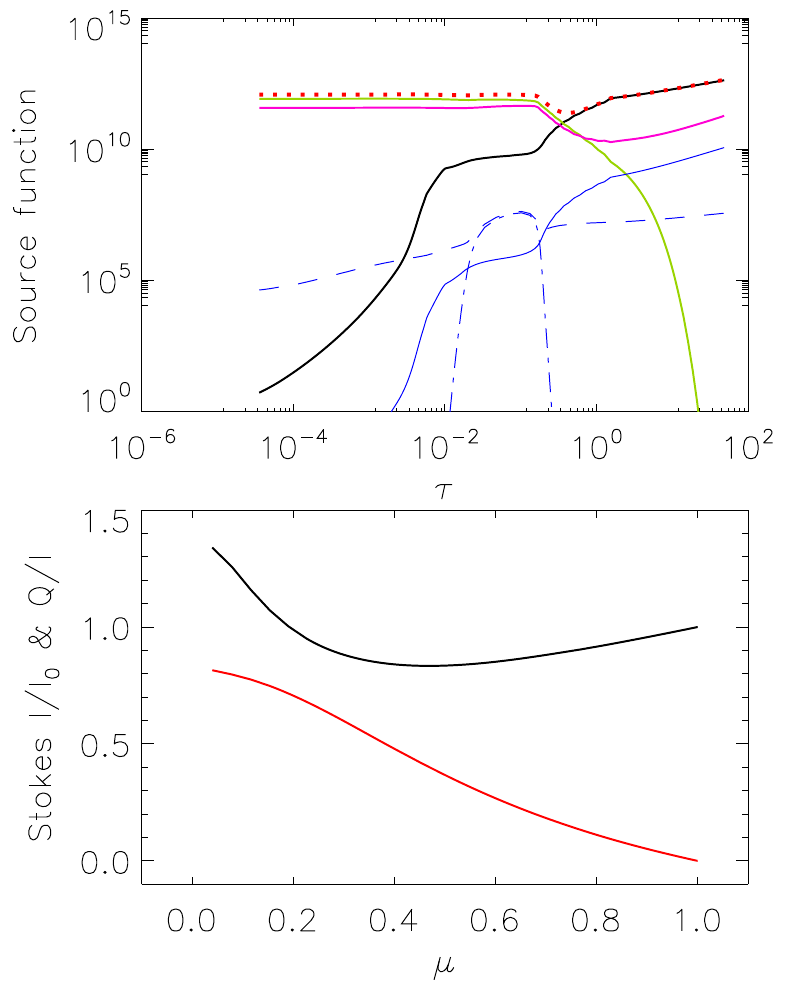}
  \includegraphics[width=0.25\paperwidth]{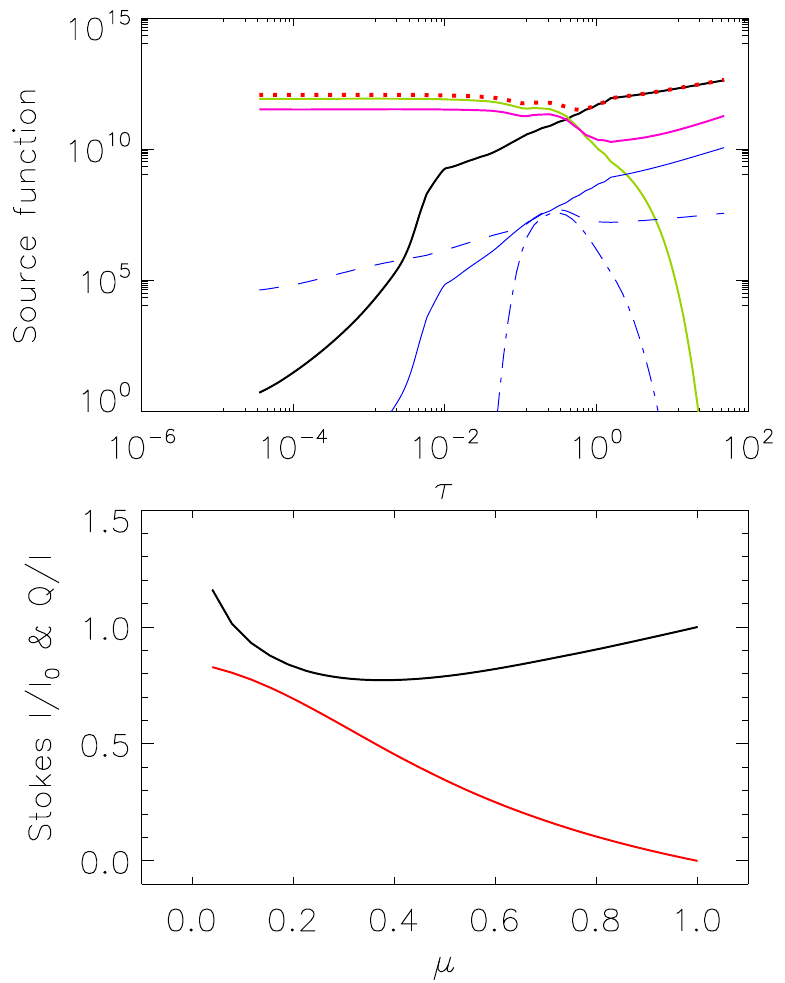}
  \caption{The same as Fig.~\ref{fig:ray} but for a planet at 0.02 AU from the star with an atmosphere
  containing {\it ad hoc} particle layers.
  The particles are assumed to have an effective size of 20nm, and the layers are at the depths of
  70, 80, and 90 km from the top of the atmosphere (plots are from left to right, respectively). 
  }
  \label{fig:cloud}
\end{figure}

Now we can model effects of particle scattering on limb intensity and polarization
distribution in planetary atmospheres by solving the polarized radiative transfer problem 
as described in Sect.~\ref{sec:prt} with the corresponding phase functions. 
As in Sect.~\ref{sec:Ray}, we investigate radiative transfer effects depending 
on irradiation and atmosphere properties. 
We use the same model atmospheres as before but replace the original layer 
of scattering particles in the planetary atmosphere with layers of various properties at different heights, 
imitating a variety of clouds. This {\it ad hoc} approach allows us 
to study intensity and polarization depending on particle (cloud) properties.
Three examples are shown in Fig.~\ref{fig:cloud}.
One can see that clouds can significantly affect the brightness 
of the irradiated atmosphere at depths (angles) 
where scattering and absorption opacities are comparable.
In the presented example of the highly irradiated planetary atmosphere 
the polarization is still determined by the single-scattered stellar photons.
In less irradiated atmospheres, the influence of particles is larger,
but still according to the scattering and absorption profiles. 
More examples with a larger variety of clouds will be published elsewhere.

\subsection{Molecular Bands}\label{sec:Mol}

Detecting molecular bands in planetary spectra is the key to their chemical composition
and to their habitability assessment.
By analysing the molecular composition we can establish whether the atmosphere is
in equilibrium or it is affected by such non-equilibrium processes like stellar
activity or life. 

Including molecular bands into polarized radiative transfer requires computation of both
line absorption and scattering coefficients. 
We compute molecular line absorption, following \cite{Berd2003}, and molecular line scattering, 
following \cite{Berd2002}, where magnetic field effects on molecular absorption and scattering 
(the Zeeman, Paschen-Back and Hanle effect) are also included and can be employed for exoplanets.
These line opacities augment the continuum opacities
at molecular band wavelengths. In addition, depending on the molecular number density 
distribution, the maximum absorption and scattering for different molecules and bands can occur 
at different heights \cite{Berd2003}. This is an important diagnostics of the atmosphere thermodynamics, 
e.g., TP profiles. 

Despite the growing amount of information, the molecular composition of exoplanetary atmospheres is 
still largely unknown. Several reported detections of molecular bands were disputed by later measurements 
(e.g., see overview and references in \cite{AframBerd2016}).
Also, a few exoplanets were found to lack any spectral features in the near infrared
which was interpreted as the presence of high clouds masking molecular absorption (e.g., \cite{Kreidberg2014}).
To explain the presence or absence of molecular bands, synthetic flux absorption and emission thermal spectra 
in clear and cloudy planetary atmospheres were computed.
Recently, it was proposed that cloud physical parameters can be constrained by a differential analysis
of various molecular bands forming at different heights in the atmosphere with respect
to the cloud height and extent \cite{AframBerd2016}. For instance, water vapor bands 
at 1.09 $\mu$m and 1.9$\mu$m show noticeably different sensitivity to particle size
and cloud extent and position at intermediate depths in the atmosphere. 
This is a sensitive spectral diagnostics of clouds.

\begin{figure}[h]
  \centering
  \includegraphics[width=0.6\paperwidth]{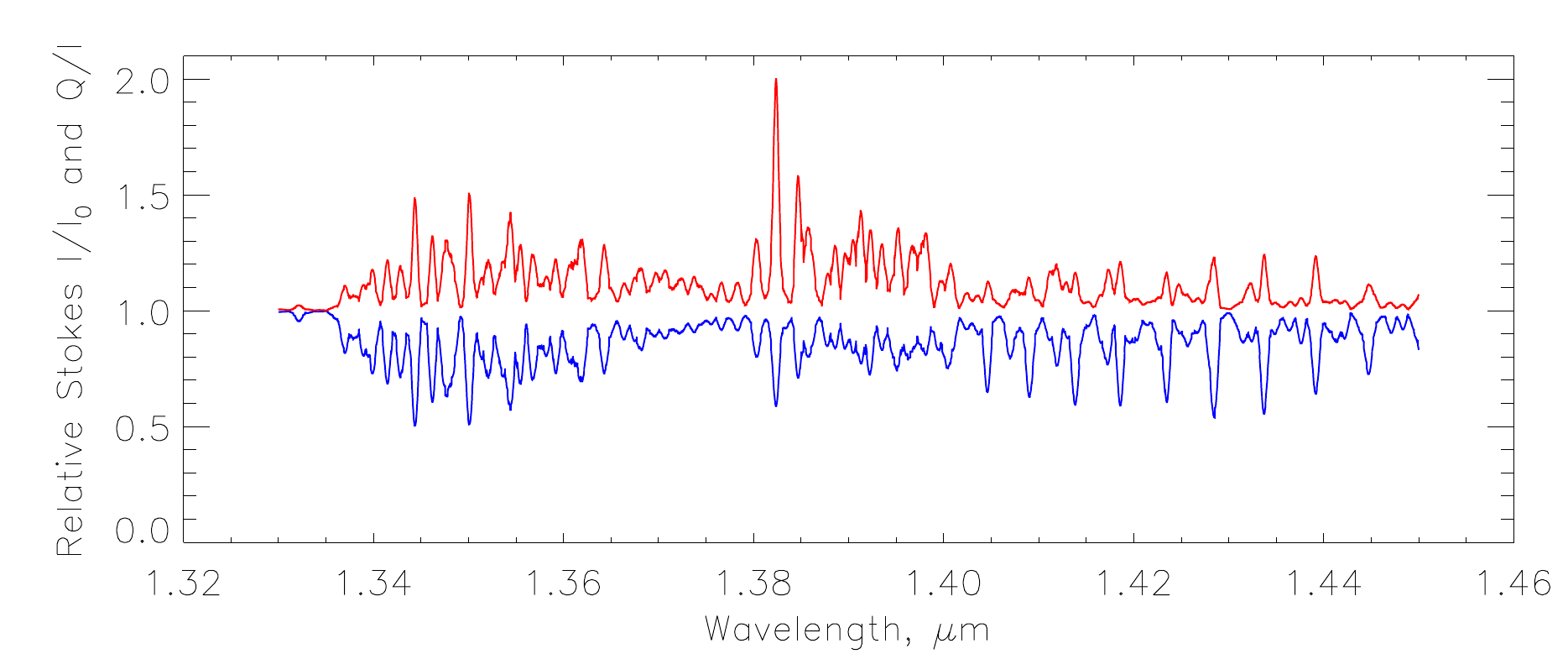}
  \caption{H$_2$O relative polarization (red) and absorption (blue) spectra plotted, respectively, up and down 
  for clarity, taking into account line polarizability. Note that there is no exact correlation between polarization 
  and absorption.
  }
  \label{fig:h2o}
\end{figure}

Polarized scattering in molecular bands was observed in the solar atmosphere and solar system planets
(e.g., \cite{Berd2002,JoosSchmid2007}). 
To model this polarization we employ the radiative transfer theory described in Sect.~\ref{sec:prt}
with the line scattering coefficient strongly dependent on wavelength (within line profiles),
line polarizability and magnetic field (if included, via the Hanle effect) \cite{Berd2002,Stenflo1994}. 
The first order radiative transfer effect leads to an apparent correlation of line scattering with absorption.
This effect increases the contrast of detection of weak signals in distant planets.
For example, model spectra from \cite{Stam2008} for the Earth atmosphere (Fig.~\ref{fig:bio}, right panels) 
show polarization in molecular oxygen and water vapour bands in red wavelengths.
However, because of the line-dependent polarizability and magnetic sensitivity, line polarization does not in general 
correlate with line absorption, as is observed in the Second solar spectrum \cite{Stenflo1994}). 
Neglecting these effects impedes the quantitative interpretation of polarization and the inferred planet parameters.
An example taking the polarizability effect into account for about 3500 H$_2$O lines near 1.4$\mu$m is shown 
in Fig.~\ref{fig:h2o}.

\subsection{Biosignatures}\label{sec:Bio}

\begin{figure}[h]
  \centering
  \includegraphics[width=0.35\paperwidth]{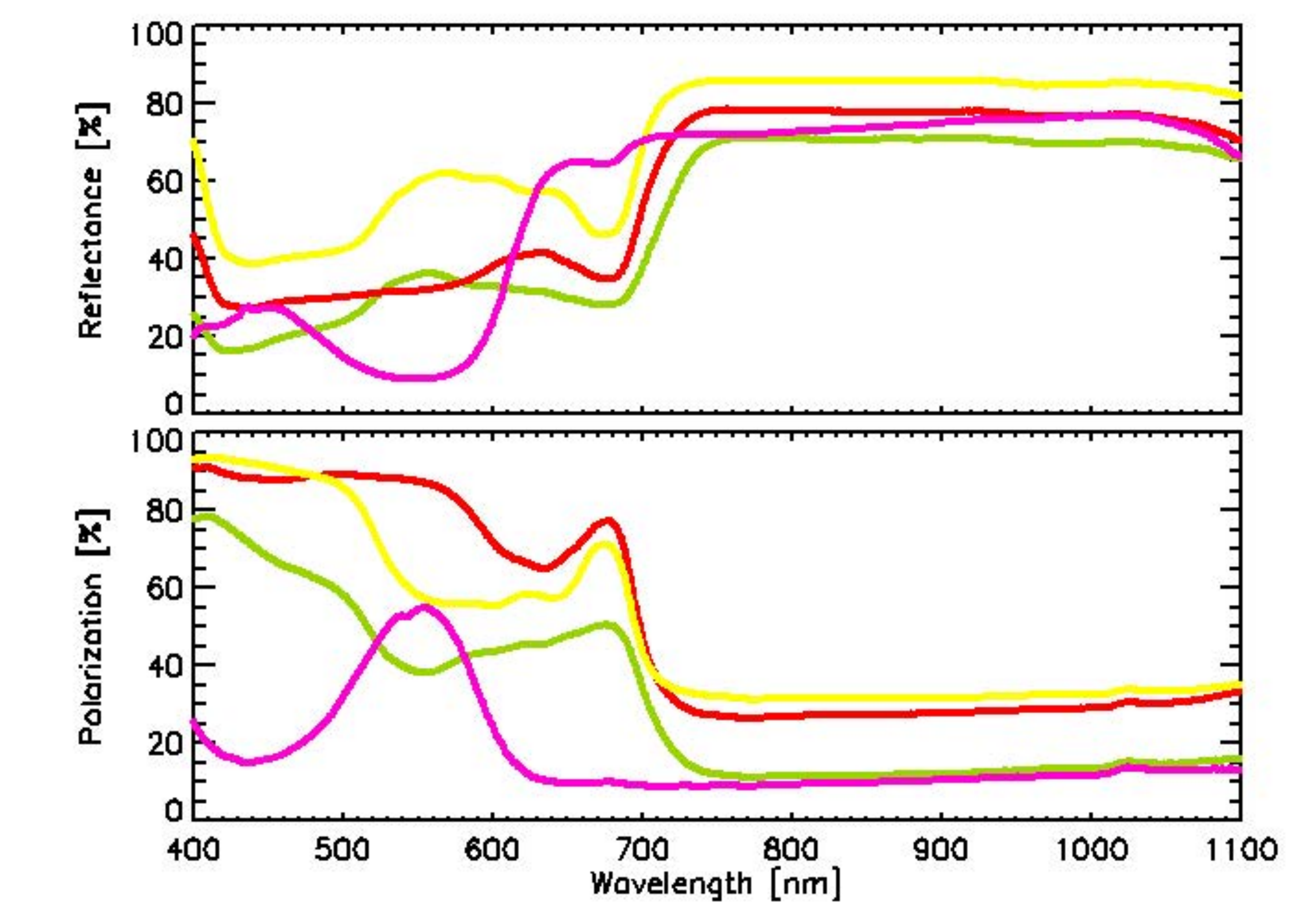}
  \includegraphics[width=0.35\paperwidth]{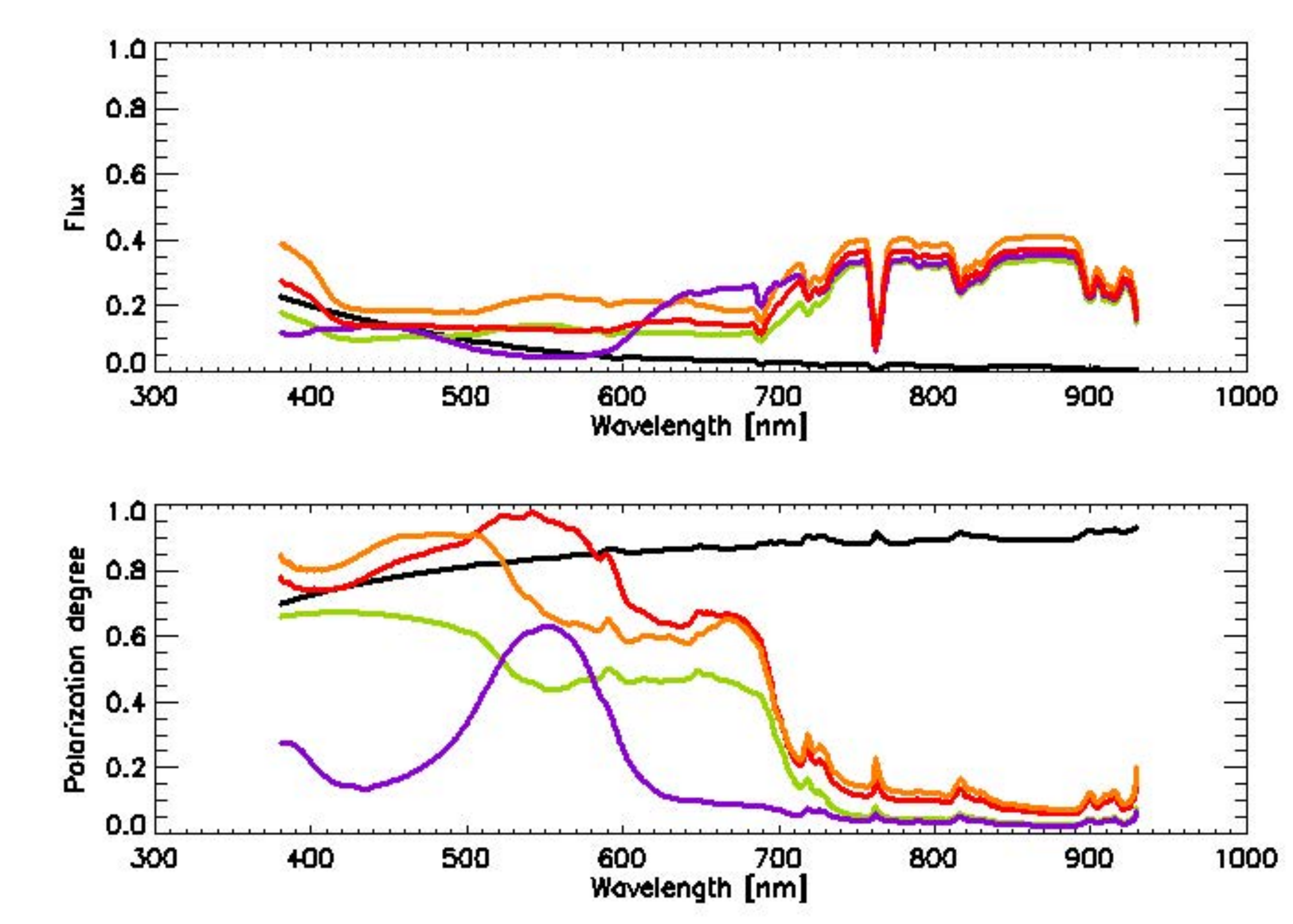}
   \caption{ {\it Left:} Reflectance and linearly polarized spectra of plant samples containing various assemblies of biopigments:
     chlorophyll (green), anthocyanins (red), carotenoids (yellow), phycobiliproteins (purple) \cite{Berd2016}. 
     Note that the higher polarization occurs at the wavelengths where these biopigments most efficiently absorb photons. 
	 The so-called ‘red edge’ near 700 nm is clearly visible. Also, polarization and reflectance are elevated 
	 if the surface of the plant is glossy (wavelength independent), cf., in the red and yellow samples.
  {\it Right:} Modelled reflectance spectra (top) and linear polarization degree spectra (bottom) 
  for planets with the Earth-like atmosphere, 80\%\ surface coverage by either of the four pigmented 
  organisms shown on the left and 20\%\ ocean surface coverage (visible hemisphere only). 
  The high linear polarization degree clearly distinguishes the presence of the biopigments in contrast to the flux spectra.
  Black curve represents a planet with an ocean only \cite{Stam2008}. The glint from its surface is highly polarized.
  }
  \label{fig:bio}
\end{figure}

A planetary surface visible through an optically thin atmosphere can be
searched remotely for spectral and polarized imprints of organisms
reflecting and absorbing stellar light.
Due to the accessibility and amount of energy provided by the stellar radiation,
it seems natural for life to evolve a photosynthetic ability
to utilize it as an energy source also on other planets.
Thus, flux spectral signatures of biological pigments
arising from photosynthesis have been
proposed as biosignatures on exoplanets \cite{Kiang2007a,Kiang2007b}. 
Moreover, it was recently shown that photosynthetic organisms absorbing 
visible stellar radiation with the help of various biopigments demonstrate 
a high degree of linear polarization associated with such absorption bands 
(see Fig.~\ref{fig:bio}). This effect was also proposed as a sensitive 
biosignature for high-contrast remote sensing of life \cite{Berd2016}.

Capturing stellar energy by photosynthetic organisms relies
on complex assemblies of biological pigments.
While chlorophyll {\it a} is the primary
pigment in cyanobacteria, algae and plants, there are
up to 200 accessory and secondary (synthesized) biopigments, 
including various forms of chlorophyll ({\it b}, {\it c} and {\it d}),
carotenoids, anthocyanins, phycobiliproteins, etc \cite{Scholes2011}. 
Various spectral sensitivity of biopigments contribute to their ability to absorb 
almost all light in the visible range (Fig.~\ref{fig:bio}).

Present and near future observations of Earth-like planets
around distant stars cannot resolve the planet surface
and image its structures directly. However, uneven
distribution of land masses and their various surface properties
as on Earth seen from space produce rotational modulation
of the reflected light which can be detected and used to
constrain the overall surface coverage of various components which can be distinguished
with flux and polarization measurements at different wavelengths.
To calculate the biosignature effect we add surface below the atmosphere
which implies new lower boundary conditions in Eq.~(\ref{eq:rt_formal}).
We allow the planetary surface to contain patches due to the presence of photosynthetic
organisms, minerals, sands and water and include also scattering and absorption in the
planetary atmosphere and clouds. 
The Earth atmosphere, ocean and clouds are the same as in \cite{Stam2008}. 
Examples are shown in Fig.~\ref{fig:bio}. 

The presence of clouds masking
the surface dilutes the information on the surface structure and composition.
A completely cloudy atmosphere will obviously disguise the presence of biopigments 
(and everything else) on the planet surface.
A small cloud coverage of around 20\%\ will only marginally reduce polarization effect 
(see \cite{Berd2016}).
Thus, clouds are the most disturbing factor in detecting
surface biosignatures, but weather variability should assist in successful
detection if a planet is monitored long enough to reveal
long-lived features on the surface.

The effect of the water ocean is also interesting \cite{Berd2016}.
The optical thickness of the ocean is basically infinite, so its
surface is dark in most colours except for the blue, where it
reflects the blue light scattered in the atmosphere.
However, there is a bright glint at the subsolar location,
which moves around the globe as the planet rotates. This
glint is due to specular reflection and is highly polarized
and practically white. Hence, an ocean only, cloud-free
planet with an Earth-like atmosphere will appear somewhat
blue (due to Rayleigh scattering in the atmosphere)
and highly polarized. 
It seems therefore that the presence of an ocean and optically thin atmosphere 
is most favourable for remote polarimetric detection of exoplanets and biopigments.

To conclude, we have presented a broad range of interesting examples where
spectropolarimetry provides novel insights into physics of exoplanets and life. 
The theoretical components outlined in this paper have been
developed since the 1950s, and they were successfully employed for probing 
atmospheres of the Earth, Sun, solar system planets, and other stars.
It is imperative now to make a full use of these techniques for advancing 
our understanding of exoplanets and for searching for life in the universe.
\\

This work was supported by the European Research Council Advanced Grant HotMol
(ERC-2011-AdG 291659). The author was fortunate to have studied at the Saint Petersburg
University under the guidance of Acad. Prof. V.V.\,Sobolev and the professors and docents 
of his faculty. The joint work in the group of Prof. Jan Stenflo at ETH Zurich
was also a great benefit for this research.



\begin{thebibliography}{9}

\bibitem{Sobolev1956}
\textit{V.V.\,Sobolev}, Radiative Transfer in Stellar and Planetary Atmospheres, Moscow, 1956.

\bibitem{Chandra1960}
\textit{S.\,Chandrasekhar}, Radiative Transfer, New York: Dover, 1960.

\bibitem{Nagirner2016}
\textit{D.I.\,Nagirner}, JQSRT, this volume, 2016.

\bibitem{FluriBerd2010}
\textit{D.M.\,Fluri, S.V.\,Berdyugina}, A\&A, \textbf{512}, A59, 2010.


\bibitem{HansenTravis1974}
\textit{J.E.\,Hansen, L.D.\,Travis}, Space Sci. Rev., \textbf{16}, 527, 1974.

\bibitem{Berd2003}
\textit{S.V.\,Berdyugina, S.K.\,Solanki, C.\,Frutiger}, A\&A, \textbf{412}, 513, 2003.

\bibitem{KostoBerd2015}
\textit{N.M.\,Kostogryz, S.V.\,Berdyugina}, A\&A, \textbf{575}, A89, 2015.

\bibitem{Kosto2016}
\textit{N.M.\,Kostogryz, I.\,Milic, S.V.\,Berdyugina, P.H.\,Hauschildt}, A\&A, \textbf{586}, A87, 2016.

\bibitem{Kosto2015}
\textit{N.M.\,Kostogryz, T.M.\,Yakobchuk, S.V.\,Berdyugina}, ApJ, \textbf{806}, 97, 2015.

\bibitem{Allard2001}
\textit{F.\,Allard, P.H.\,Hauschildt, D.R.\,Alexander, A.\,Tamanai, A.\,Schweitzer}, ApJ, \textbf{556}, 357, 2001.

\bibitem{Witte2009}
\textit{S.\,Witte, C.\,Helling, P.\,Hauschildt}, A\&A, \textbf{506}, 1367, 2009.

\bibitem{Berd2008}
\textit{S.V.\,Berdyugina, A.V.\,Berdyugin, D.M.\,Fluri, V.\,Piirola}, 
ApJ Lett., 673, L83, 2008.

\bibitem{Berd2011}
\textit{S.V.\,Berdyugina, A.V.\,Berdyugin, D.M.\,Fluri, V.\,Piirola}, 
ApJ Lett., 728, L6, 2011.

\bibitem{BerdSPW2011}
\textit{S.V.\,Berdyugina}, 
in Solar Polarization 6, eds. J.R. Kuhn et al., ASP Conf. Ser., 437, 219, 2011.

\bibitem{HansenHovenier1971}
\textit{J.E.\,Hansen, J.W.\,Hovenier}, J. Atmos. Sci., \textbf{31}, 1137, 1971.

\bibitem{vandeHulst1957}
\textit{H.C.\,van de Hulst}, Light Scattering by Small Particles, New York, Wiley, 1957.

\bibitem{Berd2003}
\textit{S.V.\,Berdyugina, C.\,Frutiger, S.K.\,Solanki}, 
A\&A, 412, 513, 2003.

\bibitem{Berd2002}
\textit{S.V.\,Berdyugina, J.O.\,Stenflo, A.\,Gandorfer}, 
A\&A, 388, 1062, 2002.

\bibitem{AframBerd2016}
\textit{N.\,Afram, S.V.\,Berdyugina}, 
A\&A, submitted, 2016.

\bibitem{Kreidberg2014}
\textit{L.\,Kreidberg, J.L.\,Bean, J.-M.\,D\'esert, et al.}, Nature, 505, 69, 2014.

\bibitem{JoosSchmid2007}
\textit{F.\,Joos, H.M.\,Schmid}, 
A\&A, 463, 1201, 2007.

\bibitem{Stenflo1994}
\textit{J.O.\,Stenflo}, Solar Magnetic Fields, Kluwer, Dordrecht, 1994.

\bibitem{Kiang2007a}
\textit{N.Y.\,Kiang, J.\,Siefert, Govindjee, R.E.\,Blankenship}, 
Astrobiol., 7, 222, 2007a.

\bibitem{Kiang2007b}
\textit{N.Y.\,Kiang, A.\,Segura, G.\,Tinetti, Govindjee, R.E.\,Blankenship, M.\,Cohen, J.\,Siefert,
D.\,Crisp, D., V.S.\,Measows}, 
Astrobiol., 7, 252, 2007b.

\bibitem{Berd2016}
\textit{S.V.\,Berdyugina, J.R.\, Kuhn, D.M.\,Harrington, T.\,Santl-Temkiv, E.J.\,Messersmith},
Int. J. Astrobiol., 15, 45, 2016.

\bibitem{Scholes2011}
\textit{G.D.\,Scholes,  G.R.\,Fleming,  A.\,Olaya-Castro, R.\,van Grondelle},
Nature Chem. 3, 763, 2011.

\bibitem{Stam2008}
\textit{D.M.\,Stam}, 
A\&A, 482, 989, 2008.





\end{thebibliography}
\end{document}